\documentclass[11pt]{article}
\usepackage{blois,epsfig}
\usepackage[latin1]{inputenc}
\usepackage[OT1]{fontenc}

\def\Journal#1#2#3#4{{#1} {\bf #2}, #3 (#4)}

\def\PRL{\em Phys. Rev. Lett.}

\def\mnras{\em MNRAS}
\def\apj{\em Astrophys. J.}
\def\aap{\em Astron. Astrophys.}
\def\araa{\em ARAA}

\def\mco{\multicolumn}

\def\ra{\rightarrow}

\def\ko{K^0}

\def\be{\begin{equation}}
\def\ee{\end{equation}}
\def\bea{\begin{eqnarray}}
\def\eea{\end{eqnarray}}

\begin{document}
\vspace*{4cm}
\title{Gamma-Ray Bursts and Particle Astrophysics}

\author{ B. GENDRE }

\address{Laboratoire d'Astrophysique de Marseille, Technopole de Marseille-Etoile, \\
38 rue Joliot-Curie 13388 Marseille CEDEX 13, FRANCE}

\maketitle\abstracts{
Gamma-ray bursts are violent events occurring randomly in the sky. In this review, I will present the fireball model, proposed to explain the phenomenon of gamma-ray bursts. This model has important consequences for the production and observation at Earth of gravitational waves, high energy neutrinos, cosmic rays and high energy photons, and the second part of this review will be focused on these aspects. A last section will briefly discuss the topic of the use of gamma-ray bursts as standard candles and possible cosmological studies.}

\section{The Gamma-ray burst phenomenon}

\begin{figure}
\begin{center}
\includegraphics[height=7cm,angle=-90]{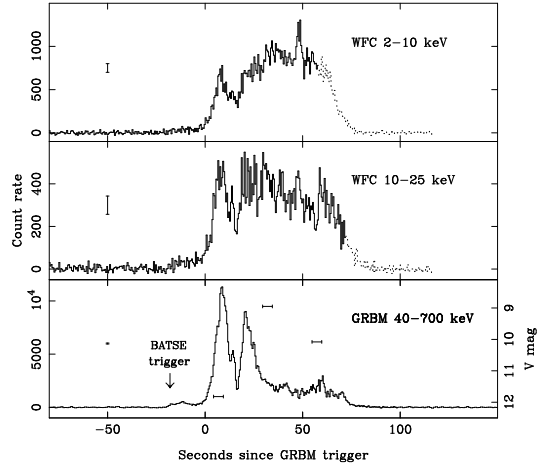}
\caption{Temporal profile of GRB 990123 at several wavelengths $^6$.
\label{grb_profile}}
\end{center}
\end{figure}

\begin{figure}
\begin{center}
\includegraphics[height=7cm, clip=true, viewport=0 0 500 350]{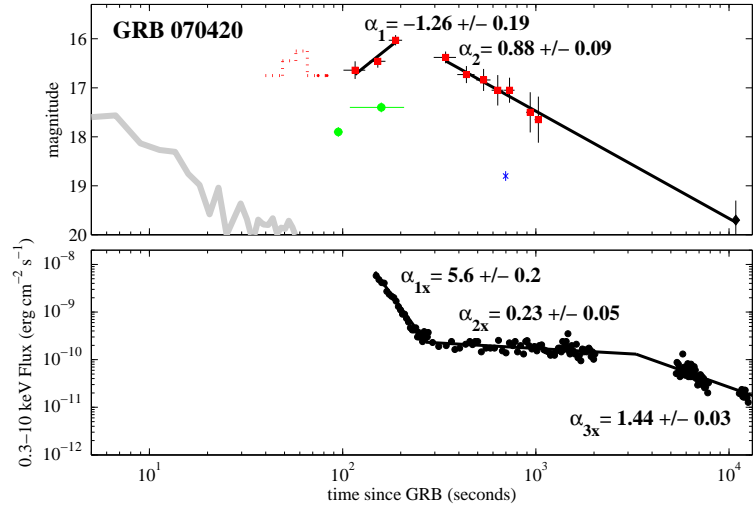}
\caption{Light curve of GRB 070420 early afterglow. Gray line corresponds to the 15-150 keV band, black points, to the 2-10 keV band, red, green and blue points to R, V, B data respectively $^7$.
\label{fig:radish}}
\end{center}
\end{figure}

Gamma-Ray Bursts (GRBs) are very powerful explosions located at cosmological distances \cite{mes06,piran04}. They are supposed to be the manifestation of newly formed black holes of few solar masses \cite{van00}. They were discovered in the late 60's by the military VELA satellites, as bursts of X-ray and gamma-ray photons occurring at random place and date in the sky \cite{kle73}. Figure \ref{grb_profile} presents a GRB temporal profile. This burst of photons, the {\em prompt} phase, lasts from few milliseconds to a few tens of seconds. It is followed by an {\em afterglow} emission emitted in all wavelengths. Depending on its brightness, this afterglow can be observed from days to months after the burst \cite{zha04}. Figure \ref{fig:radish} presents a complete light curve of GRB 070420 observed in optical, X-ray and gamma-ray bands.

There are two populations of GRBs within the Universe : {\em long} GRBs, and {\em short} GRBs \cite{dez92,kou93}. The prompt phase of long bursts lasts typically 20 seconds, and more than two seconds. The short bursts have a shorter duration, typically of 0.2 seconds. Both classes present an afterglow emission, rather similar \cite{short}. The difference between the prompt phases of short and long GRBs can be explained by a difference in the nature of the progenitor that produces the black hole. For long bursts, the black hole is the result of the core-collapse of a massive star \cite{woo93,pak98,fry99}, while for short bursts, it is the result of the merging of two compact objects \cite{pak86,goo86,ech90}. The difference in the temporal profile is due to the difference in compactness of these two progenitors.

\begin{figure}
\begin{center}
\includegraphics[height=8cm,angle=90]{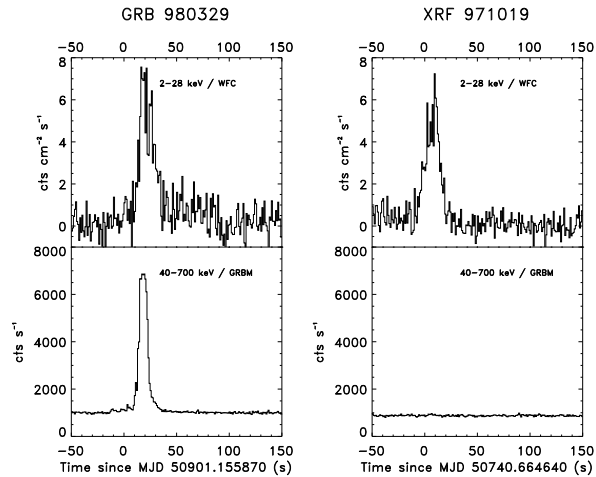}
\caption{Temporal profile of a normal GRB (left) and an XRF (right) in X-ray (top panels) and gamma-rays (bottom panels).
\label{grb_xrf}}
\end{center}
\end{figure}

GRBs are also classified on the basis of other properties. One can for example consider the optical brightness of the afterglow, distinguishing {\em dark} bursts, having no optical afterglow emission \cite{dep03}; or {\em X-ray Flashes} (which have no emission in the gamma-rays, the prompt being reduced to a burst of soft X-ray photons, see Figure \ref{grb_xrf}) \cite{hei00}. These classifications are mainly observational and can be related to the properties of the medium in which the burst develops or to the geometry of the explosion. The exact explanations of these different kinds of events are beyond the scope of this review, the interested reader can refer to De Pasquale et al. \cite{dep03}, Gendre et al. \cite{gen07} and references within these papers.

\section{The fireball model}

The newly formed black hole (and its accretion disk) is only the central engine of the fireball, that produces the observed phenomenon. The GRB is due to a jet of material traveling toward the observer (us) \cite{ree92,mes97,pan98}. The jet is accelerated up to ultra-relativistic velocity ($\Gamma > \sim 100$) by the central engine. It is not sure whether this energy injection is immediate or can last for long time \cite{inje1,inje2}. The latest SWIFT observations have shown that the central engine can sometimes be re-activated at very late time leading to late energy injection within the fireball \cite{zha06,nou06,pan06}.

Because of the relativistic beaming, the GRB phenomenon can be observed only if we are located within the jet aperture cone. A strongly off-axis position would prevent the detection of the prompt GRBs, and we could only see (in case of a long GRB) the late afterglow and a peculiar kind of supernova, an hypernova \cite{dal05}.

The jet is produced very nearby to the central engine, and a rapid estimation shows that the fireball is optically thick at that stage, and will be composed of $e^-/e^+$ pairs, photons and baryons. The baryon load of the fireball cannot be too large, otherwise the expansion would be sub-relativistic. However, most of the energy contained within the fireball will be transferred to the particles and changed into kinematic energy : this energy will further-on fuel the observed prompt and afterglow emissions where it will be reconverted into radiations during shocks.

The flow of the fireball is not emitted at a constant rate by the central engine. Even a simple assumption about chaotic physics at play within an accretion disk around a black hole implies that the fireball will be emitted by "blobs", called {\it shells}. During the acceleration process, some shells are emitted with a larger velocity compared to other ones. This differential velocity makes faster shells to catch up with slower ones emitted at earlier times. An inelastic shock will form during this interaction, accelerating electrons and producting an emission at high energy. This mechanism is called an {\em internal} shock (respect to the location within the jet), and the sum of all these interactions will produce the prompt emission \cite{ree92}.

The fireball, still expanding, will start to sweep up the surrounding material. At a given radius, the amount of matter in front of the jet is large enough to start decelerating it : at this stage a {\em forward} shock develops, that will produce the afterglow emission \cite{mes97}. The energy of accelerated electrons is distributed according to a power-law, with index $p$. The shock will generate a magnetic field within the top of the jet, and the population of electrons will start to radiate by synchrotron effect. It is common to denote the energy carried by the electrons and the energy located in the magnetic field as fractions of the total energy of the fireball ($E$). These fractions are noted $\epsilon_e$ and $\epsilon_B$ for the electron energy and the magnetic energy respectively. By definition, each of these parameters is less than one, and its sum can be (at the maximum) 1. 

\begin{figure}
\begin{center}
\includegraphics[height=7cm,angle=-90]{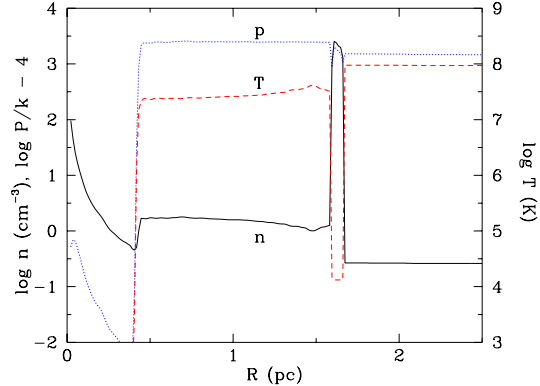}
\caption{Density profile (solid line), temperature (dashed line) and pression (doted line) of the gaz surrounding the stellar progenitor of a long GRB. From $^{32}$.
\label{profil}}
\end{center}
\end{figure}

A last parameter of the fireball is the surrounding medium density. As this surrounding medium triggers the shock producing the afterglow, its density enters within the definition of the "efficiency" of the shock to radiate its energy. This parameter, $n$, can vary with the distance, in order to take into account possible variations of the medium \cite{mes98}. With stellar progenitors, this can be expected, as massive stars produce a strong stellar wind (with a constant ejection rate, thus a decreasing density as $r^{-2}$), that will interact with the surrounding interstellar medium. This can lead to a complex density profile that can set its imprint into the afterglow light curve (see Figure \ref{profil}) \cite{che00,ram01,che04}. All the spectral and temporal variations of the afterglow are thus linked to a set of five parameters ($\epsilon_e$, $\epsilon_B$, $n$, $E$, $p$) \cite{sar98,sar99}.

\section{Implications for particle astrophysics}
\subsection{Gravitational waves}\label{subsec:prod}

By order of appearance, the first non-photonic signal that can appear for a gamma-ray burst is a signal of gravitational waves (GWs). GWs are not produced by the burst itself (i.e. the jet producing the internal and external shocks), but by the progenitor and the central engine \cite{kob03}. In the case of short GRBs, thought to be due to the merging of two compact objects, GWs can also appear before the burst : such binaries are known to radiate GWs according to general relativity. GWs can also be emitted during the production of the central black hole, either in case of a compact object merger, or in case of core-collapse. In such a case, the GW signal should arrive soon before the burst. The expected signal should be located in the range $10^2-10^3$ Hz and can be detected only for nearby events (typically a few tens of Mpc) \cite{mes06}. This low distance is the most critical problem for the detection of the gravitational signal from GRB progenitors. Under the assumption of optimal orientation, for a long GRB located at distance of 40 Mpc (as GRB 980425, the nearest GRB detected so far \cite{dista}), current GW detectors, such as LIGO or VIRGO, at nominal sensitivity will only start reaching the level of theoretical upper-limit estimates for GW emission by GRB long progenitors \cite{cor08}. The future LISA experiment, while more sensitive in other bands, should not be sensitive to the frequency band expected from the central engine GWs \cite{lisa0}: a future detection of GWs from GRBs should be done by the Advanced LIGO/VIRGO. To date, no detection of GWs from GRBs has been claimed \cite{cor08}. 

\subsection{High energy neutrinos}\label{subsec:fig}

Neutrinos can be produced in two ways within a GRB\cite{bas08}. First, any stellar progenitor will produce a flow of low energy neutrinos (few MeV) \cite{mes06}. Due to the cosmological distances of the phenomenon, such neutrinos, when produced by a GRB, cannot be detected on Earth. A second, more interesting possibility is the production of high energy neutrinos ($10^{14}-10^{17}$ eV) during the shocks leading to the prompt and afterglow emissions \cite{wax97}.

High energy neutrinos can be produced by proton-photon and proton-proton interactions. Theses kind of interactions can be numerous within the fireball. In the rest frame, the produced neutrino has a typical energy of $\sim 50$ MeV. Because of the boost due to the high Lorentz factor of the fireball, neutrinos of much larger energy are emitted \cite{wax00}. However, the exact shape of the neutrino spectrum is not known yet and strongly depends on the energy spectrum of the protons and on the energy lost by interactions made by the intermediate pion during the proton interactions. To date, the resolution and low sensitivity of neutrino experiments have not led to detection of neutrinos from GRBs.

\subsection{Cosmic rays}

The last non-photonic signal that can be sent by a GRB is a burst of protons. Indeed, the protons can be accelerated up to $10^{20}$ eV within the fireball \cite{vie95,wax95}. GRBs could thus be a possible source of ultra high energy cosmic rays (UHECR). These events have still a non clear origin, as the "GZK effect" (the interaction between the high energy protons with the microwave background photons) limits the propagation distance of such particles to less than 100 Mpc for $E_p > 10^{20}$ eV \cite{g__68,zk_68}. This should disfavor the GRBs as source of UHECR. However, because of the large number of GRBs possibly occurring within the local Universe but not observed due to off-axis observations, some of these protons may survive longer and be detected on Earth \cite{piran04}.

There cannot be a direct association between a GRB and such a cosmic ray signal. This is first due to the fact that protons are deflected by the intergalactic magnetic field, thus randomizing the position they appear on the sky \cite{piran04}. A second point is that protons are not traveling at the speed of the light. Thus, there will be a large delay between the detection of a GRB and the accelerated protons. Measurement of the cosmic ray spectrum around and above the GZK cut-off by AUGER and other experiments may strongly constrain the acceleration mechanism at play within the fireball.

\subsection{High energy gamma-ray photons}\label{sec:plac}

High energy gamma-ray photons have been observed for several GRBs. GRB 940217 produced a 18 GeV photon detected by EGRET \cite{hur94}. GRB 970417 was associated with a possible signal detected by MILAGRO in the TeV range \cite{atk00}.

High energy photons can be produced by inverse Compton scattering during the prompt and the afterglow phases \cite{pap97}. The final energy of these photons can be as large as several GeV during the prompt emission, and even more in the following afterglow emission \cite{pee04}. One may then expect a correlation between the signal seen at low energy (gamma-ray, X-ray), with the one observed at very high energy \cite{gal07}.

\section{Cosmology with GRBs}

GRBs are the most powerful explosions within the Universe since the Big Bang. They have permitted the direct detection of a single exploding star up to large distances (the record holding to date is $z = 6.3$ \cite{gen07b}). It is thus tempting to use them as cosmological probes. In the following, I will restrict my review to probes of distance estimation and direct measurement of cosmological parameters. However, it is also possible to use GRBs to study the foreground matter, and to deduce the epoch of re-ionization, the content of non-radiating matter in the line of sight, the history of the star formation rate. The interested reader can refer to Loeb \& Barkana \cite{loe00} for a more complete review on these topics.

\begin{figure}
\begin{center}
\includegraphics[height=6cm]{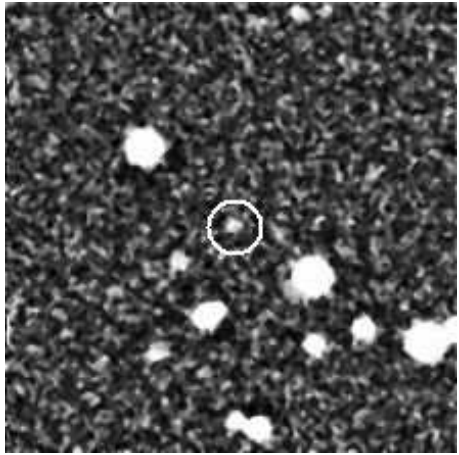}
\caption{Optical image of GRB 050904 obtained by the TAROT automated telescope at Calern (France). This small aperture telescope (25 cm diameter) caught on the fly an event occurring at z = 6.3. From $^{51}$.
\label{050904}}
\end{center}
\end{figure}

\subsection{Cosmology based on prompt observations}

Several empirical relations between observable properties of the prompt emission have been discovered during the last 10 years. Within them, the so-called Amati relation is the most discussed \cite{ama02}. It relies on a correlation between the peak of the prompt spectrum in a $\nu~f_\nu$ plane and the total energy emitted during the prompt phase assuming the isotropy. The nature of this relation is not clear. Several refinements were done on that relation, that led to the Ghirlanda relation, which takes into account the non-isotropy at the expense of a model dependent correction \cite{ghi04}. This relation allowed to draw for GRBs a Hubble diagram and thus to set constraints on the $\Omega_m - \Omega_\Lambda$ parameters \cite{ghi06}. This use of GRBs is still under debate, as the nature of the relation is not firmly established : the definition of the relation implies the use of a cosmology model (assumed to be a flat universe with $\Omega_\Lambda \sim 0.7$) which is then confirmed by the relation itself, leading to the so-called circularity problem. This problem is due to the lack of a sample of nearby GRB whose measurable properties would not depend on the cosmology. Moreover, it has been proposed that the Amati/Ghirlanda relation is a natural consequence of the fact that the total energy emitted by the fireball is computed by using a spectral model which uses the value of the peak energy \cite{but07}.

A second use of these relations is the estimation of the distance. A first attempt has been done in 2003, which gave redshift estimations not always accurate \cite{att03}. This method was refined latter, and is now accurate up to $\sim 75$ \% \cite{pel06}. This estimation, today not useful because of the limitations of the current GRB detectors in space (SWIFT/BAT has a narrow band in gamma-ray that prevents to make an accurate measurement of the value of the peak energy for most of the bursts) \cite{but07}, may become very important for the next generation of GRB detectors if the follow-up fails to measure the redshift.

\subsection{Cosmology based on afterglow observations}

Thank's to SWIFT, it is now clear that the afterglow light curves cluster around canonical afterglow light curves in X-ray \cite{gen05}, optical \cite{nar06,lia06,kan06} and near-infrared \cite{gen09}. This fact implies that the parameters of the fireball have very narrow distributions around few possible values \cite{gen08}. This fact, while not understood yet, allows measuring the redshift of bursts using their afterglow light curves, without spectroscopic observations \cite{gen06}. The method is still not very precise (with an accuracy of $\sim 50$ \%), but is a very promising tool : as soon as the origin of the clustering will be known, the precision of the method will be refined to a few percents. This will provide a very large sample of redshifts (and thus a very large sample of events) for other cosmological studies with the use of moderate diameter (1-2 meters) telescopes devoted to afterglow observations.

\section{Conclusion}

The science of GRBs is still very active. Open questions are related to the nature of the central engine, the acceleration mechanism at play during the first stages of the burst, the properties of the progenitors that lead to a GRB rather than a simple supernova, ...

France, in collaboration with China and Italy is deeply involved in a future satellite, {\it SVOM} \cite{svom0}. SVOM will be devoted to the study of the prompt and early afterglow phases, thus observing directly the last part of the acceleration stage of the fireball. Used with non-photonic instruments still in construction at the moment (such as Antares, IceCUBE or the Advanced VIRGO/LIGO) and very high energy experiments (AUGER, GLAST), this will provide very interesting constraints on these problems, ... and should of course raise new unexpected questions.

\section*{Acknowledgments}
I would like to thank the organizers for their nice invitation to this pleasant conference, and Jean-Luc Atteia, Alessandra Galli, Alessandra Corsi, Stephane Basa and Alain Mazure for comments on this review. This work was funded by the French CNES under a post-doctoral grant.

\end{document}